\documentstyle[12pt]{article}

\newcommand{\be}{\begin{equation}}
\newcommand{\ee}{\end{equation}}

\def\psnormal{\textwidth=16cm\textheight=22cm
          \oddsidemargin=0.5cm\evensidemargin=0cm
          \topmargin=0cm\parindent=1cm}
\psnormal

\begin{document}
\pagestyle{empty}

\hspace{3cm}

\vspace{-2.0cm}
\rightline{{ CERN--TH.6436/92}}
\rightline{{ IEM--FT--53/92}}

\vspace{1.0cm}
\begin{center}
{\bf SUPERSYMMETRY BREAKING AND DETERMINATION OF THE UNIFICATION
GAUGE COUPLING CONSTANT \\ IN STRING THEORIES}
\vspace{1.1cm}

B. de CARLOS${}^*$, J.A. CASAS${}^{**,*}\;$ and C. MU\~NOZ${}^{**}$
\vspace{1.3cm}

${}^{*}$ Instituto de Estructura de la Materia (CSIC),\\
Serrano 123, 28006--Madrid, Spain
\vspace{0.5cm}

${}^{**}$ CERN, CH--1211 Geneva 23, Switzerland
\vspace{1.0cm}

\end{center}

\centerline{\bf Abstract}
\vspace{0.5cm}

\noindent
We study in a systematic and modular invariant way gaugino
condensation in the hidden sector as a potential source of
hierarchical supersymmetry breaking and a non--trivial potential
for the dilaton $S$ whose real part corresponds
to the tree level gauge coupling constant (${\rm Re}\ S\sim g_{gut}^{-2}$).
For the case of pure Yang--Mills condensation, we show that no
realistic results (in particular no reasonable values for ${\rm Re}\ S$)
can emerge, even if the hidden gauge group is not simple. However, in
the presence of hidden matter (i.e. the most frequent case) there arises a
very interesting class of scenarios with two or more hidden
condensing groups for which
the dilaton dynamically acquires a reasonable value (${\rm Re}\ S\sim 2$)
and supersymmetry is broken at the correct scale ($m_{3/2}\sim
10^3\ GeV$) with no need of fine--tuning. Actually, good values for
${\rm Re}\ S$ and $m_{3/2}$ are correlated. We make an exhaustive
classification of the working possibilities. Remarkably, the results are
basically independent from the value
of $\delta^{GS}$ (the contributions from the Green--Schwarz mechanism).
The radius of the compactified space also acquires an expectation value,
breaking duality spontaneously.

\vspace{0.7cm}
\begin{flushleft}
{CERN--TH.6436/92} \\
{IEM--FT--53/92} \\
{March 1992}
\end{flushleft}
\psnormal
\psnormal
\newpage
\mbox{}

\newpage
\pagestyle{plain}
\pagenumbering{arabic}
\section{Introduction}
Recent LEP measurements corroborate [1] within a high level of
accuracy the expectations of supersymmetric grand unification [2].
Namely, the gauge coupling constants of the standard model seem
to be unified at $M_{gut}\sim 10^{16}\ GeV$ with a value $\alpha_{gut}
=g_{gut}^2/4\pi\sim 1/26$.
This unification arises naturally in the context of superstring
theories, where all the couplings are essentially the same at
tree level, even in the absence of a grand unification group. Nevertheless
there remains the fundamental question of why $g_{gut}$
should have any particular value. (Note that this problem cannot even be
posed in the framework of an ordinary GUT, where $g_{gut}$ is an input
parameter.) If superstring theories are the fundamental theory
from which the standard model is derived as a low energy limit,
they should be able to give an answer to the previous question. In fact
this is mandatory since in superstring theories we do not have
the freedom to introduce the fundamental constants as input
parameters. In particular the tree level gauge coupling constant
corresponds to the real part of a complex scalar field, the dilaton
$S$ (Re $S\equiv S_R\sim g_{gut}^{-2}$), so experimental data demand
$S_R\sim 2$. Therefore the initial
question translates as how is the vacuum expectation value (VEV)
of $S$ determined. Unfortunately $S$ has a flat scalar potential
at all order in string perturbation theory [3], at least while supersymmetry
remains unbroken. This also applies to the moduli fields $T_i$,
which are singlet under gauge interactions and whose VEVs are
usually associated with the size and shape of the compactified
space. Actually one does not have even the choice of assuming
that $<S>$, $<T_i>$ are given at present by their own cosmological
evolution since the strong bounds on the variation of
the gauge coupling constants with time and on scalar--tensor
theories of gravity exclude this possibility [4]. Therefore,
non--perturbative effects must play a fundamental role in the
determination of $<S>$, $<T_i>$.

As has been suggested above this process should be strongly
related to the supersymmetry breaking mechanism. This is by
itself one of the most important technical problems which
remains unsolved in superstring theories. More precisely, it is
not yet understood how supersymmetry can break at a scale
hierarchically smaller than the Planck scale $M_P$, which is
essential for correct low energy physics. The most promising
source of non--perturbative effects capable of generating this
hierarchical supersymmetry breaking is gaugino condensation in the
hidden sector [5].
The reason for this is that the scale of gaugino condensation
corresponds to the scale at which the gauge coupling becomes
large and this is governed by the running of the coupling
constant. Since the running is only logarithmically dependent on
the scale, the gaugino condensation scale is suppressed relative
to the initial one by an exponentially small factor
$\sim e^{-\frac{1}{2\beta g^2}}$ providing a promising source for the
mass hierarchy ($\beta$ is the one--loop coefficient of
the beta function\footnote{
$\beta=\frac{1}{16\pi^2}(3C(G)-\sum_R T(R))$, where
$C(G)$ is the Casimir operator in the adjoint representation of $G$,
$R$ labels the chiral supermultiplets transforming like representations
of $G$ and $T(R)$ is defined by Tr$(T^iT^j)=T(R)\delta^{ij}$.}
of the hidden sector gauge group $G$). However in
string theories it is not obvious this mechanism will work for,
as mentioned above, the value of the gauge coupling at the
starting scale for evolution is itself a dynamical variable. It
should be mentioned here that these are effects in the low
energy field theory. At present there is no framework for
calculating non--perturbative effects in the full string theory,
so one has to assume that field theory non--perturbative effects
dominate over the stringy ones. On the other hand this seems
reasonable since we are examining the physics at low
energy, where quantum field theory is a good
approximation of string theory.

Much work has recently been devoted to the
study of gaugino condensation in the framework of string
theories [6--14]. It has been learned that the assumption of
duality, i.e. target
space modular $SL(2,Z)$ symmetry in the effective four dimensional
supergravity action [15], modifies the usual form of
the condensate [8--10] in essential
agreement with previous one--loop calculations
including string heavy modes [16]. This modification, which has
been slightly corrected in ref.[14] by taking into account the
contributions from the Green--Schwarz mechanism [17,18], has proved to be
extremely useful to fix the VEVs of some of the moduli
$T_i$ [8,9]. It has also been shown that no further
modifications arise from higher string loops [19]. The influence of
hidden matter, which is usually present, in the condensation
process has also been studied [7,11--14,20]. However, in spite of all these
achievements there are no known scenarios working
satisfactorily up to now. In particular, reasonable values for the
dilaton (and thus for the gauge coupling constant) have proved to be
very difficult to obtain. It has been suggested that the existence
of two or more
condensates in the hidden sector [21,6,7,22], as well as the presence of
hidden matter [7,12], may contribute to fixing the dilaton correctly
and breaking supersymmetry at the right scale. However, no
systematic analysis of these possibilities has been performed
yet. This is precisely the main goal of this paper.

Our starting point is the effective field theory coming from
a compactified string theory leaving one supersymmetry intact. The
resulting $D = 4$ $N = 1$ supergravity theory [23] is characterized
by the superpotential $W$ and the gauge kinetic function $f$,
holomorphic in the chiral superfields, and the K\"{a}hler
potential $K$, real--analytic in the chiral superfields. These
functions are determined, in principle, in a given
compactification scheme, although in practice they are known
completely only for orbifold compactification schemes [24], which
on the other hand have proved to possess very attractive features
from the phenomenological point of view [25]. More precisely, in the
general case when the gauge group contains several factors $G =
\prod_{a} G_{a}$, the exact gauge kinetic functions in string
perturbation theory, up to small field--independent contributions,
are [16--18]
\begin{eqnarray}
f_W^a = k^aS + \frac{1}{4\pi^2}\sum_{i=1}^3 (\frac{1}{2}b_i^{'a} - k^a
\delta_{i}^{GS}) \log(\eta(T_{i}))^{2}\;\;,
\label{fw}
\end{eqnarray}
with
\begin{eqnarray}
b_{i}^{'a} = C(G^a)-\sum_Q T(R_Q^a)(1+2n_Q^i)\;\;,
\label{bp}
\end{eqnarray}
where the meaning of the various quantities appearing in
(\ref{fw},\ref{bp}) is the following:
$k^{a}$ is the Kac--Moody level of the $G^{a}$ group ($k^a=1$ is
a very common possibility),
$T_{i}$ ($i = 1, 2, 3$) are untwisted moduli, whose real parts
give the radii of the three compact complex dimensions of the
orbifold (Re $T_{i} = R_{i}^{2}, i = 1, 2, 3$),
$\delta_{i}^{GS}$ are 1--loop contributions coming from the
Green--Schwarz mechanism, which have been
determined for the simplest (2,2) $Z_{N}$ orbifolds [17],
$\eta(T_{i})$ is the Dedekind function, $Q$ labels the matter fields
transforming as $R^a_Q$ representations under $G^a$ and $n_Q^i$ are the
corresponding modular weights (see below).
The 1--loop K\"{a}hler
potential, up to terms involving matter fields, is [17]
\begin{eqnarray}
K_{1-loop} & = & -\log( S + \bar S + \frac{1}{4\pi^2}\sum_{i=1}^3
\delta_{i}^{GS}
\log(T_{i} + \bar{T_{i}})) - \sum_{i=1}^3 \log(T_{i} + \bar{T_{i}})
\nonumber \\
           & = & -\log Y - \sum_{i=1}^3 \log(T_{i} + \bar{T_{i}})\;\;.
\label{K}
\end{eqnarray}
The function $Y=S + \bar{S} + \frac{1}{4\pi^2}\sum_{i} \delta_{i}^{GS}
\log(T_{i} + \bar{T_{i}})$ can be considered [17] as the redefined
gauge coupling
constant at the unifying string scale: $Y=\frac{2}{g^2_{string}}$.
Finally, the perturbative superpotential $W^{pert}$ has the form
\begin{eqnarray}
W^{pert} = h_{IJK} Q^{(u)}_{I}Q^{(u)}_{J}Q^{(u)}_{K} +
h^{'}_{IJK}(T_{i}) Q^{(t)}_{I}Q^{(t)}_{J}Q^{(t)}_{K} +
h^{''}_{IJK} Q^{(u)}_{I}Q^{(t)}_{J}Q^{(t)}_{K} + \cdots\;\;,
\label{Wpert}
\end{eqnarray}
where $Q^{(u)}_{I}$ $(Q^{(t)}_{I})$ are untwisted (twisted) charged
matter fields.
The value of $h_{IJK}$, $h^{''}_{IJK}$ for the allowed couplings
is simply one, while $h^{'}_{IJK}(T_{i})$ are complicated but
known functions of $T_{i}$ [26,27]. The dots in (4) stand for
non--renormalizable terms and other terms involving twisted
moduli which are not relevant for the discussion here. We will
make use of the previous expressions throughout the paper.

As mentioned above the theory is invariant under $[SL(2,Z)]^3$
target space modular transformations, which are of the form
\begin{eqnarray}
T_{i} \hspace{0.4cm} \rightarrow \hspace{0.4cm}
\frac{a_iT_{i}-ib_i}{ic_iT_{i}+d_i} , \hspace{0.4cm} a_id_i-b_ic_i = 1,
\hspace{0.4cm} a_i, b_i, c_i, d_i \in Z\;\;.
\label{modular}
\end{eqnarray}
This means in particular that the combination
${\cal G} \equiv K+\log |W|^{2}$
must be modular invariant which, using (\ref{K}), implies that
the superpotential has to transform with modular weight $-3$ [15], i.e.
\begin{eqnarray}
W \hspace{0.5cm}   \rightarrow \hspace{0.5cm} W \prod_{i=1}^3
(ic_iT_{i}+d_i)^{-1} \;\;\; ,
\label{modW}
\end{eqnarray}
and the dilaton $S$ has to transform as [17]
\begin{eqnarray}
S \hspace{0.5cm} \rightarrow \hspace{0.5cm} S+\frac{1}{4\pi^2}\sum_{i=1}^3
\delta_{i}^{GS} \log(ic_iT_{i}+d_i)\;\; .
\label{modS}
\end{eqnarray}
Note that the function $Y$ defined in (\ref{K}) is modular invariant.
Similarly, charged matter fields $Q$ transform as
\begin{eqnarray}
Q \hspace{0.5cm}   \rightarrow \hspace{0.5cm} Q \prod_{i=1}^3
(ic_iT_{i}+d_i)^{n_Q^i} \;\;\; ,
\label{modQ}
\end{eqnarray}
In particular, untwisted matter fields associated with the $j$--th
complex plane have modular weights $n^i_Q=-\delta^i_j$. Modular
weights corresponding to various types of twisted matter fields can
be found in ref.[28].
The dependence of $h_{IJK}$, $h'_{IJK}$, $h''_{IJK}$ on $T_{i}$
is such that $W^{pert}$ has modular weight $-3$.

In section 2 the case of pure Yang--Mills gaugino condensation
in the hidden sector is analysed in a systematic way for one, two
or more condensing subsectors taking the $\delta^{GS}=0$ case as
a guide example and extending afterwards the results to $\delta^{GS}\neq
0$. We show that no realistic results (in particular no reasonable
values for $S_R$) can emerge in this context, even if
the the hidden sector gauge group is not simple.
In section 3 we consider the most frequent case,
namely gaugino condensation in the presence of matter. We again
take $\delta^{GS}=0$ and matter of the untwisted type as a useful
guide example, eventually extending the results
to the most general case, i.e. $\delta^{GS}\neq 0$ and matter in
the twisted sector. In this way there arises a
wide and very interesting class of scenarios with two hidden
condensing groups $SU(N_1)\times SU(N_2)$ for which
the dilaton dynamically acquires a reasonable value ($S_R\sim 2$)
and supersymmetry is broken at the correct scale ($m_{3/2}\sim
10^3\ GeV$). Actually, good values for ${\rm Re}\ S$ and $m_{3/2}$
turn out to be correlated. We have made an exhaustive analysis of the working
possibilities (see tables), including also Kac--Moody levels
$k\neq 1$ and other scenarios with more than two
condensing groups. In section 4 we present our conclusions.

\newpage
\section{Gaugino condensation without matter}
\subsection{General aspects}

Examples of pure Yang--Mills (YM) hidden sectors in superstring
constructions can be found in the literature. For instance, the
$E_8'$ gauge group for the standard embedding of $Z_N$ and
$Z_N\times Z_M$ orbifolds is well known. Also $Z_7$, $Z_4$ models
with $SU(3)\times SU(2)\times U(1)^5$ observable sector and
pure $E_6$, $SU(3)\times SU(3)$ gauge hidden sectors respectively
were constructed [29] and many more examples can be built up.

The process of gaugino condensation in the context of a pure
Yang--Mills (YM) $N = 1$ supergravity theory has been pretty well
understood for a long time [5]. It can be conveniently
described [9] by an effective superpotential $W^{np}(U)$ of the
chiral composite superfield $U$ ($= \delta_{ab} W^{a}_{\alpha}
\epsilon^{\alpha\beta} W^{b}_{\beta}$) whose scalar component
corresponds to the gaugino composite bilinear field
($\lambda\lambda$). $W^{np}$ reads
\begin{eqnarray}
W^{np} = \frac{U}{4} (f_{W} + \frac{2}{3} \beta
\log U)\;\; .
\label{Wnp}
\end{eqnarray}
In our case $f_{W}$ is given by eq.(\ref{fw}) with $b_i'=16\pi^2 \beta/3$.
As it has been demonstrated in refs.[7,11,12], it is equivalent to work
with (\ref{Wnp}) than with the resulting superpotential after
substituting $\partial W^{np}/\partial U = 0$, i.e.
\begin{eqnarray}
W^{np} = d\ e^{-\frac{3}{2\beta}f_{W}} =
d \ \frac{e^{-\frac{3k}{2\beta}S}}{\prod_{i=1}^3[
\eta(T_i)]^{(2- \frac{3k}{4\pi^2\beta} \delta_i^{GS})} }\;\;,
\label{super}
\end{eqnarray}
where $d = -\beta/6e$. Notice that in the previous expression
the form of $f_W$ is valid at all order in perturbation theory [19].
If the gauge group is not simple,
$G = \prod_{a} G_{a}$, then $W^{np} = \sum_{a}
W_{a}^{np}$. Using the K\"ahler potential of eq.(\ref{K}), the scalar
potential (valid for a generic $W(S,T)$) is given by [14]
\begin{eqnarray}
V & = & \frac{1}{Y\prod_{i=1}^{3}(T_i+\bar T_i)}
 \{  |YW_{S}-W|^{2} + \sum_{i=1}^3 \frac{Y}{Y+\frac{1}{4\pi^2}
\delta^{GS}_i}|(W+\frac{1}{4\pi^2}\delta^{GS}_iW_S)
\nonumber \\
  &   &  - (T_i+\bar T_i)W_{T_i}|^2 - 3|W|^2 \}\;\;,
\label{V}
\end{eqnarray}
where $W_\phi=\partial W/\partial \phi$, $\phi=S,T_i$ and $Y$ is defined
in eq.(\ref{K}).
The simplest case occurs when $\delta_i^{GS}=0$. This happens for
instance in the $Z_{2} \times Z_{2}$ orbifold with standard
embedding [17]. From now on we will take this as our
guide example since it actually gives the essential features of all
the cases, but we will address in subsection 2.5 all the (slight)
departures from the $\delta_i^{GS}=0$ results in a generic case.
Notice that for $\delta_i^{GS}=0\;$ $W^{np}$ admits
the decomposition $W^{np}(S,T_i) = \Omega(S)/\prod_{i=1}^3
\eta(T_i)^{2}$, even if
$G$ is not simple. For the sake of
simplicity we will work for the moment with just an overall
modulus $T=T_1=T_2=T_3$, parametrizing the global size of the compactified
space. Then the scalar potential is given by [8]
\begin{eqnarray}
V = \frac{| \eta(T) |^{-12}}{2S_{R}(2T_{R})^{3}} \{
|2S_{R}\Omega_{S}-\Omega |^{2} + (\frac{3T_R^2}{\pi^2}|\hat{G}_2|^2
-3) | \Omega |^{2} \}\;\;,
\label{potencial}
\end{eqnarray}
where $\hat{G}_{2} = -(\frac{\pi}{T_{R}} + 4\pi\eta^{-1}
\frac{\partial \eta}{\partial T})$ and
$S_R\equiv {\rm Re}\ S$, $T_R\equiv {\rm Re}\ T$. As has been noted
in ref.[8], for a given reasonable value of $S$ this
potential always develops a minimum in $T$ very close to the
point $T = 1.2$.
Therefore, {\em provided $S$ takes a
reasonable value}, the theory will be automatically compactified
with a very sensible value of $T$. However, it is not obvious
that the former will happen as will become clear shortly.
The condition of stationary point in the dilaton, i.e.
$\partial V/ \partial S = 0$, can be satisfied in two different ways:
\begin{eqnarray}
    \hspace{1cm} & i)  & \hspace{1cm}2S_{R} W_{S}-W = 0
\label{mincon1} \\
    \hspace{1cm} & ii) & \hspace{1cm}4S_{R}^{2}\Omega_{SS}
(2S_{R}\Omega_{S}-\Omega)^{*} = E\Omega^{*}(2S_{R}\Omega_{S}-\Omega)
\label{mincon2}
\end{eqnarray}
with $E = 2-\frac{3T_{R}^{2}}{\pi^{2}}|\hat{G}_{2}|^{2}$.
In order to see whether {\em i)} and {\em ii)} correspond to
true minima in $S$ or not it is necessary to calculate the
Hessian\footnote{This was performed in ref.[8] for $\Omega(S) =
c+de^{-3S/2 \beta}$ and $c<<1$, finding that condition (\ref{mincon1})
always corresponds to a
minimum in $S$ in the weak coupling limit ($S_{R}\stackrel{>}{\sim}1$).}.
Working
in the general case $\Omega(S) = \sum_{a} d_{a}
e^{-3k_{a}S/2 \beta_{a}}$ it is convenient to define the parameters
$x_{a} = 2S_{R}(3k_a/2 \beta_{a})$, which for any reasonable value
of $S_{R}$ and $\beta_{a}$ are much larger than 1. For the first
possibility {\em i)} it can be verified that the Hessian $H = V_{pp}
V_{qq}-V_{pq}^{2}$ (with $p =$Re$\ S = S_{R}$, $q =$Im$\ S$) can be
expressed as a sum of several terms of
different order in $x_{a}$. Retaining only the dominant part we find
\begin{eqnarray}
V_{pp} = V_{qq} = \frac{4S_R}{|\eta(T)|^{12}(2T_R)^3}
|\Omega_{SS}|^2
\label{derpi}
\end{eqnarray}

\begin{eqnarray}
V_{pq} = \frac{-2}{|\eta(T)|^{12}(2T_R)^3}
\ {\rm Im}(\Omega_{SS}\Omega_{S}^{*})\;\; .
\label{derpqi}
\end{eqnarray}
So $|V_{pq}| << |V_{pp}|$ and $H>0$. Consequently, if (\ref{mincon1})
has a solution for reasonable $S_{R}$,
it will correspond to a minimum. Analogously, for the case {\em
ii)}, retaining again the dominant part in $x_{a}$, we find
\begin{eqnarray}
V_{pp} = -V_{qq} = \frac{-8S_R^2}{|\eta(T)|^{12}(2T_R)^3 E}
\ {\rm Re}(\Omega_{SSS} \Omega_{SS} e^{-i \sigma})
\label{derp}
\end{eqnarray}

\begin{eqnarray}
V_{pq} = \frac{-8S_R^2}{|\eta(T)|^{12}(2T_R)^3 E}
\ {\rm Im}(\Omega_{SSS} \Omega_{SS} e^{-i \sigma})\;\;,
\label{derpq}
\end{eqnarray}
where $(2S_{R} \Omega_{S} - \Omega) = \mid 2S_{R} \Omega_{S} -
\Omega \mid e^{i \sigma /2}$. Therefore $H<0$ and we conclude
that the case (\ref{mincon2}) never corresponds to a realistic minimum.
We have checked these results numerically.

Hence, it is enough in the subsequent analysis to focus our
attention on the solutions of type (\ref{mincon1}). It is also
worth noticing that these solutions
always lead to the same values of $T$ in the minimum,
independently of the value of $S$, as it is easily deduced from
(\ref{potencial},\ref{mincon1}); more precisely $T=1.23$.
Of course other minima, with
the same physical characteristics, appear for all the values of
$T$ related to the previous one by a modular transformation
(\ref{modular}).

\subsection{Hidden sector with a unique condensate}
This case has already been considered in the literature.
Inserting simply (\ref{super}) (with $\delta^{GS} = 0$) in
(\ref{potencial}) we find a potential without minima in $S$,
as it can be deduced from (\ref{mincon1}).
This is schematically shown in fig.1 where $V$ is
represented versus $S_{R}$ for a typical group and setting $T$
at its minimizing value. The maximum in $S_{R}$ corresponds to
condition (\ref{mincon2}). Therefore, depending on the initial condition for
$S_{R}$, it either runs away to infinity, thus leading to a
free theory, or enters in a strong coupling regime
($S_{R}\rightarrow 0$). Both cases are clearly unrealistic. For
the general case ($\delta^{GS} \neq 0$) the situation is
exactly the same (see subsection 2.5 below).

\subsection{Hidden sector with two gaugino condensates}
The possibility of having two or more gauge group factors in the
hidden sector has been previously considered in refs.[21,6,7,12,22].
This is in fact the most usual scenario in four dimensional strings
(see e.g.[30]). However the minimization of the complete
dual--invariant potential has not been performed up to now.
Before starting it should be said that in string theory the
rank of the hidden gauge group $G=\prod G_a$ is strongly restricted
by the fact that its contribution to the total conformal central
charge, plus that of the standard model, cannot exceed 22.
Hence, $c_{hidden}=\sum_a c_a=\sum_a \frac{dim(G_a)k_a}
{k_a+C(G_a)}\leq 22-c_{SM}$, where $c_{SM}$ is the contribution
of the standard model. For $k=1$, $c$ coincides with the rank of the
corresponding group, otherwise it is higher. Consequently rank$(G)\leq 18$,
where the equality can only be achieved if all the groups, including
the standard model one, are level 1. This restriction is crucial
to imagine possible scenarios. Let
us consider, to start with, a compactification in which the
hidden sector is pure YM with gauge group $G = G_{1} \times G_{2}$,
where $G_{1}$ and $G_{2}$ are simple, and $\delta^{GS}=0$. Then the
corresponding condensation superpotential, see eq.(\ref{super}), is

\begin{eqnarray}
W^{np} = d_{1} \frac{e^{-3k_{1}S/2\beta_{1}}}{\eta^{6}(T)} + d_{2}
\frac{e^{-3k_{2}S/2\beta_{2}}}{\eta^{6}(T)}\;\;,
\label{supdos}
\end{eqnarray}
where $\beta_{1}$, $\beta_{2}$ are
the 1--loop coefficients of the $G_{1}$ and $G_{2}$ beta
functions respectively. As we know from section 2.1 any
realistic minimum of the potential $V(S, T)$, see
eq.(\ref{potencial}), arising from (\ref{supdos}) should
satisfy the condition (\ref{mincon1}), which in this case reads:

\begin{eqnarray}
2S_{R}W_{S}-W = 0 \hspace{0.5cm} & \rightarrow & \hspace{0.5cm}
S_{R} =  \frac{2}{3(k_{1} \beta_{1}^{-1}-k_{2} \beta_{2}^{-1})}
\log{\frac{d_{1}(3S_{R}k_{1} \beta_{1}^{-1}+1)}
{d_{2}(3S_{R}k_{2} \beta_{2}^{-1}+1)}}
\label{sr}\\
\hspace{0.5cm} & \rightarrow & \hspace{0.5cm}
{\rm Im}\ S  = \frac{2\pi(2n+1)}{3(k_{1} \beta_{1}^{-1}-k_{2}
\beta_{2}^{-1})} \ , n \in Z\;\;.
\label{ims}
\end{eqnarray}
Clearly the value of $S$ is not related to the specific
$T$--dependence of $W$.
Notice that Im$\ S$ is simply such that the two
condensates get opposite signs giving a cancellation between
them. If $S_{R}$ is realistic, i.e. $S_{R} = O(1)$, then
$3k_aS_{R}\beta_{a}^{-1} >> 1$ and condition (\ref{sr}) becomes
analogous to
\begin{eqnarray}
W_{S} = 0 \hspace{0.5cm} \rightarrow \hspace{0.5cm} S_{R} =
\frac{2}{3(k_{1} \beta_{1}^{-1}-k_{2} \beta_{2}^{-1})}
\log{\frac{d_{1}\beta_{2}k_{1}}{d_{2}\beta_{1}k_{2}}}\;\; .
\label{Ws}
\end{eqnarray}
The condition $W_{S} = 0$ was used in refs.[7,12] to look for
realistic scenarios. We see here that it can be properly
justified. Clearly the value of $S_{R}$ is related to the value
of the $d_{a}$ coefficients. Using the field theory value
$d_{a} = -\beta_{a}/6e$, see eq.(\ref{super}), it can be easily
verified that eq.(\ref{sr}) cannot be fulfilled for positive
$S_{R}$ and $k_{1}=k_{2}$. For $k_1\neq k_2$ it is possible to find
non--trivial solutions to eqs.(\ref{sr},\ref{ims}), but they
are always far from the realistic range, i.e. $S_R<<1$.
E.g. for $SU(3)_{k_1=1}\times SU(4)_{k_2=2}$ and
$SU(6)_{k_1=1}\times SU(7)_{k_2=2}$ there appear minima at
$S_R=0.0447$ and $S_R=0.0535$ respectively. These results cannot
be essentially improved for other choices of the gauge group.
Certainly $T$ becomes fixed at a reasonable value, but the fact
that the gauge coupling constant enters in the strong coupling regime
does not allow this result to be taken very seriously.

It has been argued in ref.[6] that a possible way out of this
problem is the appearance of "stringy" field--independent
threshold corrections $\Delta_{a}$ to be added to eq.(\ref{fw}). Then
the value of $d$ in eq.(\ref{super}) is modified in a different
way for each condensate
\begin{eqnarray}
d_{a} \hspace{0.4cm} \rightarrow \hspace{0.4cm} d'_{a} = d_{a}
e^{-3 \Delta_{a}/2\beta_{a}}\;\;.
\label{di}
\end{eqnarray}
Depending on the size of these corrections, it
might occur that eq.(\ref{sr}) has a reasonable solution for
some choice of $G_{1}$ and $G_{2}$. On the other hand, these
contributions have been calculated in ref.[16] in the context of
the $Z_{3}$ orbifold and in ref.[31] for fermionic constructions
associated with $Z_{N} \times Z_{M}$ compactifications. In all
cases they are very small, far from fixing up the
situation. So we conclude that this possibility is rather
speculative, a reasonable choice being to take $\Delta_{a}=0$,
as we will do throughout. In any case {\em all} the results of
the paper are straightforwardly modified using eq.(\ref{di}) to
count the effect of these field--independent contributions.

\subsection{Hidden sector with three or more condensates}

The previous results are not improved when more than
two condensates are considered. In general, if the hidden sector
gauge group contains $n$ factors $G = \prod_{a=1, \ldots
,n}G_{a}$, then
\begin{eqnarray}
W^{np} = \sum_{a=1, \ldots ,n} d_{a}
\frac{e^{-3k_{a}S/2\beta_{a}}}{\eta^{6}(T)}
\label{Wnp2}
\end{eqnarray}
with the same notation as in eq.(\ref{supdos}). As was shown in
subsection 2.1 condition (\ref{mincon1}) is the only one that can lead to
realistic minima. This implies
\begin{eqnarray}
\sum_{a=1}^{n} z_{a} = 0\;\;,
\label{sumz}
\end{eqnarray}
where $z_{a} = d_{a} (1+3S_{R}k_{a}\beta^{-1}_{a})
e^{-3k_{a}S/2\beta_{a}}$. For $n=3$, i.e. in the case of three
condensates, (\ref{sumz}) leads to $|z_{a}+z_{b}| = |z_{c}|$ ($a \neq
b \neq c \neq a$) which, denoting $z_{a} = |z_{a}|
e^{i\varphi_{a}}$, reads
\begin{eqnarray}
\cos(\varphi_{a}-\varphi_{b}) = \frac{|z_{c}|^{2}  -
|z_{a}|^{2}-|z_{b}|^{2} }{
2|z_{a}|\ |z_{b}| },\;\;\;a \neq b \neq c \neq a \;\;.
\label{cosfi}
\end{eqnarray}
Now, using $-1 \leq
\cos(\varphi_{a} - \varphi_{b}) \leq 1$, we obtain bounds on
$S_{R}$ which turn out to be surprisingly strong. In fact they
virtually exclude any possibility of finding realistic minima
with three condensates, as is shown in table 1. (Actually
it is even difficult to find examples
with true minima, as the numerical analysis reveals.) These
results are basically maintained for $n > 3$. In ref.[22] a
stationary point was found for the rather bizarre case $G =
SU(2)_{k=1} \times SU(2)_{k=2}^{4} \times SU(2)_{k=3}^{4}$, where
$k$ denotes the levels of the associated Kac--Moody algebras. We
have found that this stationary point really corresponds to a
minimum, but it occurs at $S_{R} = 0.017$, again a
completely unrealistic value. In addition this solution does not
break supersymmetry.

\subsection{The $\delta^{GS} \neq 0$ case}

The general case occurs when $\delta_i^{GS} \neq 0$. This is reflected in
eqs.(\ref{K},\ref{super}) and in the resulting potential
$V$, given by eq.(\ref{V}). The complexity of the latter does not
allow for an analytical study as that performed for $\delta^{GS}_i=0$.
In particular, conditions (\ref{mincon1},\ref{mincon2}) now become
much more involved.
However, the numerical analysis shows that $V$
essentially exhibits the same features as
the one for the $\delta_i^{GS} = 0$ case, see eq.(\ref{potencial}).
Namely, for a unique condensate there is no minimum at all
in $S$, while for two or more condensates examples can be found with
minima but for unrealistic values of $Y=2/g_{string}^2$. In any case, it is
interesting to notice that it may be impossible for the last scenario
(i.e. two or more pure YM sectors) to occur in many instances,
as modular invariance arguments indicate. To see this consider
first the $Z_{3}$ and $Z_{7}$ cases. These orbifolds do not possess
$N=2$ subsectors, which implies that the coefficient of $\log
\eta(T_{i})$ in eq.(1) must vanish [16,17], i.e. $b^{'a}_{i} =
16\pi^2\beta^a/3=2k^a\delta^{GS}_i$. Taking into account
that $\delta_{i}^{GS}$ are universal for all the gauge group
factors it is clear that this cancellation cannot take place
simultaneously for two sectors with different $(k^{-1}\beta)$
coefficients. Alternatively, notice that, since for these orbifolds
$\eta(T)$ does
not appear in the expression of the condensation superpotential (\ref{super}),
there is no modular transformation of $S$ (see eq.(\ref{modS})) able
to give the correct modular weight $-3$ to a sum of pure YM
condensates $W^{np} \propto \sum_{a} e^{-3k_aS/2 \beta_{a}}$
unless all the $(k^{-1}_a\beta_{a})$ factors coincide, which for
practical purposes is equivalent to having a unique condensate.
This argument can be
extended to the rest of the $Z_{N}$ orbifolds since all of them
have planes which are never completely rotated. Then the coefficient
of $\log \eta(T_{i})$ in
eq.(\ref{fw}) for the associated modulus must vanish again [16,17], leading to
the same conclusion as for the $Z_{3}$ and $Z_{7}$ orbifolds.
This rule is actually satisfied by all the pure YM examples constructed
up to now, e.g. the above mentioned $SU(3)\times SU(3)$ $Z_4$ model
of ref.[29].

On
the contrary, there is nothing in principle against having
multiple pure YM gaugino condensation in $Z_{N} \times Z_{M}$
orbifolds.

\vspace{1cm}
The results of this section show that a scenario with a pure YM
hidden sector is not viable from the phenomenological point of
view. The most interesting situation takes place when there are
two or more condensates, which, apparently, is only possible for
$Z_{N} \times Z_{M}$ orbifolds. However, the minima of the
scalar potential found in this case are clearly unrealistic, since
they have $S_R<<1$. This leads us to consider in the next section
the most general (and frequent) case, namely gaugino condensation
in the presence of matter.

\section{Gaugino condensation in the presence of matter}
\subsection{General procedure and the $\delta^{GS}=0$ case}

It is known that the existence of hidden matter is the most general
situation and occurs in all promising string constructions [25,32].
For orbifold compactifications, explicit condensing
examples can be found. E.g. in ref.[25] phenomenologically
interesting $Z_3$ models with $G=SU(3)\times SU(2)\times U(1)_Y
\times [SO(10)]'$ and three 16's hidden matter representations
were constructed. In ref.[33] two $Z_7$ models with
$SU(3)\times SU(2)\times U(1)^5$ observable gauge group and
$SO(10)$, $SU(5)\times SU(3)$ hidden sectors with hidden matter
10, $3(5+\bar 5)+7(3+\bar 3)$ respectively were constructed.
The rules for building up $[SU(3)\times SU(2)\times U(1)^5]\times
[G]'$ models in the framework of $Z_3$ and $Z_7$ orbifolds are
given in [34,33]. Examples in the framework of other $Z_N$ orbifolds,
e.g. a $Z_4$ model with $E_6\times SU(2)$ observable sector and $E_7$
hidden gauge group plus $2(56)$ hidden matter representations,
can be found in [30,35]. Many more examples can be constructed.

Gaugino condensation in the presence of {\em massive} matter
fields was studied several years ago for globally
supersymmetric field theories [36--38]. However, there is at present
no generally accepted formalism describing the condensation in the presence of
massless matter (see refs.[37,38]) and it should be noticed that, strictly
speaking, in the context of superstring theories there are no
light matter fields with mass terms in the $D=4$ effective
Lagrangian. Fortunately, these fields usually have trilinear
couplings between them (see eq.(4)) and, therefore, one can
proceed as if they were massive with a dynamical mass given by
the VEV of another matter field. Furthermore one has to take into
account that the effective field theory is now locally supersymmetric.
This has been done in the past
year in refs.[11,12,14]. We will just write here the final result. In
the case of $G = SU(N)$ with $M(N + \bar{N})$ "quark"
representations  $Q_{\alpha}$, $\bar{Q}_{\alpha}$, $
\alpha=1,\cdots,M$,
coupled to a set of singlet fields $\{A_r\}$ in the
usual way (see eq.(\ref{Wpert}))
\begin{eqnarray}
W^{tr} = \sum_{r,\alpha,\beta}h_{r\alpha\beta}(T_i) A_r
Q_{\alpha} \bar{Q}_{\beta}
\label{trilineal}
\end{eqnarray}
(of course many Yukawa couplings $h_{r\alpha\beta}$ can be zero),
the complete condensation superpotential can be written as
\begin{eqnarray}
W^{np}(S,T,A) = -N(32 \pi^{2} e)^{\frac{M}{N}-1}
[\det {\cal M}]^{\frac{1}{N}}\ \frac{e^{-\frac{8 \pi^{2}k}{N}
S}}{\prod_{i=1}^3[\eta(T_i)]^{\frac{2}{N}(b_i'-2k\delta^{GS}_i)}}\;\;,
\label{cond}
\end{eqnarray}
where ${\cal M}_{\alpha\beta}=\sum_{r}h_{r\alpha\beta}(T_i) A_r$.
In this result the trilinear piece $W^{tr}$
(eq.(\ref{trilineal})) has also been incorporated. It can be checked
that (\ref{cond}) has the correct modular weight $-3$. Notice that
the exponent in the numerator is simply $-3kS/2 \beta$ where $\beta$
corresponds
to the pure YM $SU(N)$ theory. $W^{np}$ in eq.(\ref{cond}) is for
quark masses ($\sim h(T_i)A$) smaller than the condensation scale
$\Lambda$. In the opposite case the result is exactly the same
but with a slightly different prefactor. The non-vanishing Yukawa
couplings $h_{r\alpha\beta}$
are simply 1 for untwisted or mixed couplings while they have a non--trivial
dependence on $T_i$ for the twisted case [26,27], see eq.(\ref{Wpert}).
These formulae are easily extended to the case
of a non--simple gauge group $G=\prod_a G_a$.

Of course, for this approach to be consistent, the VEVs of the
$A$ fields must become at the end different from zero as a consequence
of the minimization of the potential,
otherwise the "quark" fields are not massive. However, this has not
been studied in the literature yet, and it is not obvious to
happen. As in the previous section we take the $\delta^{GS}_i=0$
case as a useful guide example (which is known to occur in the $Z_2\times
Z_2$ orbifold with standard embedding [17]) and in the next
subsection we will extend the results to the
general case. We also work for convenience with an overall modulus
$T$, a generic $A$ field of the untwisted
type (so $\det {\cal M}=A^{M}$) and untwisted $Q$ fields\footnote{
It is worth noticing here that, concerning modular transformations
and Yukawa couplings, most of the matter fields in $Z_{N} \times
Z_{M}$ orbifolds "behave" as untwisted ones.
For $Z_N$ examples with all the hidden matter in the untwisted sector
see e.g. refs.[30,35].}.
This simplified approach gives
the correct results when the number of quarks to which $A_i$ ($i$ denotes
a holomorphic index) gives mass in (\ref{trilineal}), say $M_i$, is
the same in the three complex planes (i.e. $M_1=M_2=M_3=\frac{1}{3}M$).
Then the minimization equations allow  $<A_1>=<A_2>=<A_3>\equiv <A>$,
$<T_1>=<T_2>=<T_3>\equiv <T>$. Under more general circumstances this
approach gives an average of $<A_i>$, $<T_i>$. The K\"ahler potential
for untwisted fields and $\delta^{GS}_i=0$ is very well known [39]
\begin{equation}
K_{1-loop} = -\log( S + \bar{S})
-3\log[(T + \bar{T}) - |A|^2]\;\;.
\label{K2}
\end{equation}
The corresponding modular invariant potential is thus
\begin{eqnarray}
V & = & \frac{1}{2S_{R}(2T_{R}-|A|^{2})^{3}} \{ \mid 2S_{R}W_{S}-W
\mid^{2} + \frac{1}{3}(2T_{R}-|A|^{2}) \mid \frac{\partial
W}{\partial A}+ \bar{A} \frac{\partial W}{\partial T} \mid^{2}
\nonumber \\
  &   & + \frac{1}{3} (2T_{R}-|A|^{2})^{2} \mid \frac{\partial
W}{\partial T}-\frac{3W}{2T_{R}-|A|^{2}} \mid^{2} - 3 \mid W
\mid^{2} \}\;\;.
\label{potmat}
\end{eqnarray}
For $|A|^{2} = 2T_{R}$ there exists an infinite barrier which allows
us to restrict the values of $|A|^{2}$ to the range
$|A|^{2}<2T_{R}$. This is also the condition for kinetic energy
positivity. Substituting (\ref{cond}) in (\ref{potmat}) for one
or more condensates corresponding to a group $G = \prod_{a=1}^n
G_{a}$, one finds a quite involved expression for $V(S,T,A)$
which requires a numerical analysis. This analysis indicates that $V$
does not have true minima for $n<3$. This behaviour can be intuitively
understood in the following way. For a realistic case ($S_{R}$, $T_{R} =
O(1)$) one expects $<|A|^2>\  << S_{R},T_{R}$, since $A$ is expected to
have a vanishing VEV at the perturbative level. At least this certainly
should happen for $|A|< \Lambda$, which, as will become clear
shortly, is the relevant case for us. Then, since
$W^{np}$ in (\ref{cond}) is a sum of terms that are
power--law functions of $A$, it is clear that the term $\propto
| \partial W/ \partial A|^{2}$ in (\ref{potmat}) dominates
the potential (except within a very narrow band around $\partial W/
\partial A = 0$). Hence, we can approximate
\begin{eqnarray}
V \sim \frac{1}{6S_{R}(2T_{R})^{2}} \mid \frac{\partial
W}{\partial A} \mid^{2}\;\;,
\label{potap}
\end{eqnarray}
which has the absolute minimum at
\begin{eqnarray}
\partial W/ \partial A = 0
\label{Wa}
\end{eqnarray}
and possible relative minima at $\partial^{2} W/
\partial A^{2} = 0$ (note that the requirement (\ref{Wa}) leaves
global SUSY unbroken in the weak coupling limit). No one of these
conditions can be non--trivially fulfilled with the superpotential of
eq.(\ref{cond}) when a single condensate is considered.
Therefore $V(S,T,A)$ does not have minima. These heuristic results
are fully confirmed by the numerical analysis. When several
condensates are involved the situation is more complex and is studied
in subsection 3.3, where it is shown that only for three or more
condensates and under particular circumstances there can emerge
interesting solutions.
%\footnote{In particular for two condensates "sharing" the
%same $A$ field in (\ref{cond}) the equation (\ref{Wa}) has solution, but
%it is not difficult to see that the resulting superpotential $W(S,T)$
%after eliminating $A$ is similar to that for a single pure YM condensate
%(see eq.(\ref{super})), so the corresponding potential does not have
%minima in $S,T$.
%We have found, however, some bizarre cases with actual minima. For example,
%for $G=SU(2)\times SU(3)\times SU(4)\times SU(5)$ with matter content
%$M=1,0,1,3$ respectively and two fields $A_1$, $A_2$ associated with
%$SU(2)\times SU(3)$ and $SU(4)
%\times SU(5)$, there is a minimum at $S_R=0.404,
%T=1.23$. In any case no realistic results emerge.}.

The situation expounded above might seem phenomenologically
discouraging. Fortunately this is not so, thanks to the crucial role
played by the perturbative superpotential. The importance of the
interplay between the perturbative and non--perturbative
parts of the superpotential for a good supersymmetry breaking
has already been pointed out in refs.[40,7] and, more recently, in
ref.[13]. The point is that, besides the coupling to the quarks,
$A$ has perturbative interactions with other singlet fields.
This changes the situation dramatically. To see this, consider
the following simplified ansatz\footnote{We thank J. Louis
for putting forward this possibility to us.} for $W$
\begin{eqnarray}
W &=& W^{np} + W^{pert}
\label{supmat}   \\
W^{pert} &=& A^{3}
\label{supert}
\end{eqnarray}
(recall that $A$ is assumed to be untwisted). Now, it is easy to check
that condition (\ref{Wa}) has a non--trivial solution. For the case of
a unique condensate eq.(\ref{Wa}) leads to
\begin{eqnarray}
A^{3} = \left(\frac{M}{3}\right)^{\frac{3N}{3N-M}}(32\pi^2 e)^
{\frac{3(M-N)}{3N-M}}\
\frac{e^{-24\pi^2kS/(3N-M)} } {\eta(T)^{6}} \;\;,
\label{Ast}
\end{eqnarray}
where we have used $\sum_i b^{'}_i=3N-M$ in the untwisted case.
Note that certainly if $S_R, T_R = O(1)$ then $<|A|^2>\  <<
S_{R},T_{R}$, as is required for consistency.
Furthermore $A^3=\frac{M}{96\pi^2}U$, where $U$ is the gaugino composite
bilinear field (see section 2), which indicates that $|A|<\Lambda$
is indeed the relevant case for us. Moreover, the F--term associated with $A$
($\propto [W_A+W K_A+\bar A(W_T+WK_T)]$) is clearly dominated by $W_A$. Hence,
 we can
substitute (\ref{Ast}) in $W$, thus obtaining an effective superpotential
that depends on $S$ and $T$ only
\begin{equation}
W^{eff} =  \tilde d\; \frac{e^{{-3kS}/{2\tilde{\beta}}} }
{\eta^6(T)}\;\;\;,
\label{Weff}
\end{equation}
\begin{equation}
\tilde {\beta} = \frac{3N-M}{16\pi^2}\;,\;\;
\tilde{d} = (\frac{M}{3}-N)(32\pi^2 e)^{\frac{3(M-N)}{3N-M}}\left(
\frac{M}{3}\right)^{\frac{M}{3N-M}}\;\;,
\label{dtilde}
\end{equation}
where $\tilde{\beta}$ is the beta function of the complete $SU(N)$
theory with quarks. Again, the numerical analysis confirms the validity
of this procedure for "integrating out" the $A$ field.
Notice that $W^{eff}$ still has the correct modular weight $-3$,
exhibiting the same form as in the pure YM case\footnote{This
was first noticed by Kaplunovsky and Louis [41,13].},
eq.(\ref{super}), except for the different values of the exponent
($-3k/2 \tilde\beta$) and the prefactor $\tilde{d}$ (this modification will
become crucial). In consequence, the form of the scalar
potential of eq.(\ref{potencial}) and the subsequent analysis
hold for this case. In particular, we can assure that, even in
the presence of matter, the potential for a unique condensate
does not develop true minima.
Things change dramatically, however, for multiple gaugino
condensation. The case of two condensates is especially
interesting. Then\footnote{We are supposing here for simplicity that
the $A$--field giving mass to the quarks of the first gauge group is
different from that of the other.}
\begin{eqnarray}
W^{eff} = \tilde{d}_{1} \frac{e^{-3k_1S/2 \tilde\beta_{1}}}{\eta^{6}(T)} +
\tilde{d}_{2} \frac{e^{-3k_2S/2 \tilde\beta_{2}}}{\eta^{6}(T)}\;\;.
\label{supeff}
\end{eqnarray}
As was shown in subsection 2.3 the associated potential can
only have a realistic minimum at
\begin{eqnarray}
2S_{R}W^{eff}_{S}-W^{eff} = 0 \hspace{0.5cm} & \rightarrow &
\hspace{0.5cm} S_{R} =
\frac{2}{3( k_1\tilde{\beta}_{1}^{-1}- k_2\tilde{\beta}_{2}^{-1})}
\log{\frac{\tilde{d}_{1}(3S_{R}k_1
\tilde{\beta}_{1}^{-1}+1)}{ \tilde{d}_{2} (3S_{R}k_2
\tilde{\beta}_{2}^{-1}+1)}}
%\nonumber \\
\label{sr2}\\
\hspace{0.5cm} & \rightarrow & \hspace{0.5cm} {\rm Im}\ S =
\frac{2 \pi(2n+1)}{3( k_1\tilde{\beta}_{1}^{-1}-
k_2\tilde{\beta}_{2}^{-1})}, n \in Z
\label{ims2}
\end{eqnarray}
(see eqs.(\ref{sr}, \ref{ims})). Recall here that this condition is
equivalent to $W_S^{eff}=0$ for $S_R=O(1)$, see eq.(\ref{Ws}).
The crucial difference with respect
to the pure YM case comes from the values of $\tilde{\beta}_{i}$,
$\tilde{d}_{i}$. These depend only on which the gauge groups and
matter contents of the hidden sector are. It will be seen shortly
that there are many possibilities for these leading to
a value of $S_{R}$ ($=g_{string}^{-2}$) in the
realistic range ($S_{R} \sim 2$). In this way, there naturally arises
a wide and very interesting class of scenarios for which a good perturbative
unification is plausible. For the sake of definiteness we have represented
in table 2 the complete set of $SU(N_1)\times SU(N_2)$ scenarios
with $k_1=k_2=1$,
$N_a>M_a$ and $N_1\neq N_2$ for which $1\leq S_R\leq 3.5$ (for $N_1=N_2$
there are many more possibilities, as is commented below). We
have allowed for this rather wide range because the gauge coupling
constant $g_{gut}^2$ at the unification mass ($M_{gut}\sim 10^{16}\ GeV$)
is not  exactly $S_R^{-1}$ but it is corrected by threshold
contributions [16,17] and the running between $M_{string}$ and $M_{gut}$
(see ref.[42]). Both facts are model--dependent. We are also assuming
that the standard model gauge group has $k=1$, otherwise the
realistic range for $S_R$ would be $1\leq k S_R\leq 3.5$,
thus leading to additional classes of viable scenarios.
Notice however that for $k\neq 1$ the range of the hidden gauge group
is restricted further. These results are in impressive agreement
with those obtained numerically by using the complete potential of
eq.(\ref{potmat}) and the superpotential of eqs.(\ref{supmat},\ref{supert}).
As in the pure YM case the value of $T$ in the
minimum is always $T = 1.23$ or any other value obtained from this by a
modular transformation (\ref{modular}), so duality is spontaneously broken
as it was noted in refs.[8,9].
Notice that the values of $S$ and $T$ are not related to each other,
i.e. the solution of $\partial V/\partial S=0$ does not depend
on the specific dependence of $K$ and $W$ on $T$ and vice versa.
We have represented the potential for a typical case in fig.2, where
duality invariance can be clearly recognized.

To understand intuitively these results it is convenient to recall
that conditions (\ref{sr2}, \ref{ims2}) for $S_R=O(1)$ are almost
equivalent to $\partial W^{eff}/\partial S =0$ (see eq.(\ref{Ws})).
If we make some further approximations, neglecting small
logarithms, we obtain a useful formula for $S_R$
\begin{equation}
S_R \simeq 0.17\ \frac{N_2M_1-N_1M_2}{k_1(3N_2-M_2)-k_2(3N_1-M_1)}\;\;,
\label{sr3}
\end{equation}
corresponding to a gauge group $SU(N_1)\times SU(N_2)$ with
$M_1(N_1+\bar N_1)$ and $M_2(N_2+\bar N_2)$ quark representations.
Clearly $S_R$ can be $O(1)$ for many combinations of $N_1,N_2,M_1,M_2$,
as is illustrated in table 2,
so we see that the requirement of a reasonable gauge coupling constant
does not imply fine--tuning. In particular, it is clear from (\ref{sr3})
that for a hidden gauge group
$SU(N)_{k=1}\times SU(N)_{k=1}$, provided $M_1\neq M_2$, the value
of $S_R$ is essentially independent from the matter content ($S_R\sim 0.17N$)
and quite reasonable for $6\leq N\leq 10$ (notice here that
$SU(10)_{k=1}\times SU(10)_{k=1}$ has the maximum rank allowed).
This fact is illustrated at the end of table 2 with a few examples
out of the 145 existing ones.  It is worth noticing that examples
with $N\leq M<3N$ can also be found (some of them
are shown in table 3). This seems to be a departure from previous
results in the context of global supersymmetry. This is in fact
a polemical point in the literature. While  it is claimed in ref.[37]
that for $N\leq M$ no effective superpotential describing gaugino
condensation can be formulated since the "squark" condensate
$\Pi= \det (\bar Q_iQ_j)$ vanishes identically, in ref.[38] it is argued
that this condensate can be a different expression of the squark fields,
thus permitting normal gaugino condensation. In this second case
the scenarios found with $N\leq M$ are on the same footing as the others.
If we allow for $k_1\neq k_2$ many additional scenarios
appear, as is reflected in table 4.
Finally, let us emphasize the fact that all the minima found previously
do correspond to a weak coupling regime, which makes these solutions
very reliable.

Once we have seen that it is possible to fix the dilaton satisfactorily
in a wide class of scenarios, let us examine the supersymmetry
breaking issue. Actually, in the previously determined minima supersymmetry
is spontaneously broken due to the non--vanishing of the F--term
associated with $T$. Notice that the $S$ F--term is zero due to the
minimum condition (\ref{sr2},\ref{ims2}). On the other
hand, the cosmological constant $\Lambda_{cosm}$ is generically different from
zero and negative, so we have to assume the existence of some unknown
contributions rendering $\Lambda_{cosm}=0$. The supersymmetry
breaking scale is essentially given by the gravitino mass $m_{3/2}$, which
for the scenarios considered above (i.e. two condensing groups in the
hidden sector) is
\begin{equation}
m_{3/2}=e^{K/2}|W^{eff}| \simeq \frac{1}{(2S_R)^{1/2}
(2T_R)^{3/2}|\eta(T)|^6}\
|\tilde{d}_{1}(1-\frac{k_1\tilde\beta_2}{k_2\tilde\beta_1})|\
e^{-3k_1S_R/2 \tilde\beta_1}\;\;,
\label{m32}
\end{equation}
where $W^{eff}$ is given by (\ref{supeff}) and we have used $W^{eff}_S=0$
which is a condition practically equivalent to (\ref{sr2}) for realistic
$S_R$. The requirement of having $10^2\ GeV\leq m_{3/2}\leq\ 10^4 GeV$
(which is necessary
in order to preserve the virtues of supersymmetry to solve the hierarchy
problem) restricts the viable possibilities of the hidden sector further.
There survive, however, many of them. In particular, for
$k_1=k_2=1$ there are 527 viable scenarios. Eleven of them have $N_a>M_a$
($\#\ 6,9,12,14,17,19,28,32,35,36,37$ in table 2). The rest have
$M\geq N$ in one of the two sectors
(80 of them are shown in table 3). Clearly, there is still no need
of fine--tuning in the hidden sector to arrange realistic values for
$g_{string}$ and $m_{3/2}$ simultaneously. All this all reflects the capacity
of gaugino condensation as the source of a hierarchical supersymmetry
breakdown.
To understand the numeric results intuitively  notice that the most
important factor in (\ref{m32}) is the exponential. Using the approximate
value of $S_R$ given in eq.(\ref{sr3}) and doing some further
approximations it can be seen that a realistic value for $m_{3/2}$
requires
\begin{equation}
M_2-M_1\simeq 6(N_2-N_1)\;\;.
\label{empeq}
\end{equation}
This empirical formula is well satisfied by the 527 above mentioned
scenarios. A very remarkable fact is that once we require a reasonable
value for $m_{3/2}$ the corresponding value of $S_R$ is automatically
placed about $S_R=2$, i.e. the optimum value (see tables 2, 3). Therefore,
good values for $g_{string}^{-2}$ and $m_{3/2}$, far from being
incompatible, are correlated! Actually this pleasing and non--trivial
result is a characteristic of level 1 constructions, which seem to be
favoured from this point of view. E.g. for $k_1=1$, $k_2=2$ there are
76 examples with $S_R$ and $m_{3/2}$ inside reasonable ranges (5 of them
with $N>M$), to be
compared with the 527 examples for $k_1=k_2=1$. They are displayed in
table 4, from which it is clear that the typical value of $S_R$ is
located about $S_R=1$.

Let us finally see that, as usual, soft breaking masses
for the scalars similar to the gravitino mass are generated.
The dependence of the K\"ahler potential and the scalar potential
on an untwisted field $Q$  is the same as that on $A$ in
eqs.(\ref{K2},\ref{potmat}). Consequently, using eq.(\ref{Wa})
and $W_S^{eff}\simeq 0$ (see eqs.(\ref{sr2},\ref{ims2})),
the physical mass
$m_Q$ generated for the canonically normalized scalar
component of $Q$ can be written, after some algebra, as
\begin{equation}
m_Q^2 = m_{3/2}^2 - e^K |\frac{1}{3}(2T_R-|A|^2)
\frac{\partial W}{\partial T}-W|^2 + V_o + \dots\;\;,
\label{mQ}
\end{equation}
where $m_{3/2}^2=e^K|W|^2$, $V_o$ is essentially the cosmological
constant (which we are assuming to be zero) and the dots stand
for other terms involving $A$, which are negligible once one takes
into account the fact that $|A|^2<<S_R,T_R$, see eq.(\ref{Ast}).
Substituting now the explicit form for $W$ of eq.(\ref{supeff}) we find
\begin{equation}
m_Q^2 \simeq \left(1-\left(\frac{T_R}{\pi}\right)^2 |\hat G_2(T)|^2\right)
m_{3/2}^2 \simeq (0.81 m_{3/2})^2\;\;,
\label{mQ2}
\end{equation}
where $\hat G_2(T)$ is defined in eq.(\ref{potencial}) and we have
used the value of $T$ in the minimum, i.e. $T=1.23$. Notice that
$m_Q$ is modular invariant.

\subsection{The general case}

The results so far obtained in this section can be extended to the
$\delta^{GS}_i\neq 0$ case with some modifications. Working still
with hidden matter of the untwisted type (recall that this is the general
situation in many $Z_N$ and $Z_N\times Z_M$ constructions)
we can proceed as in the previous subsection in order to eliminate the
$A$ field through condition\footnote{To reproduce step by step the
arguments of subsection 3.1 it is necessary to use the K\"ahler potential
$K$ for untwisted fields and $\delta^{GS}\neq 0$, which is not very
well known although modular invariance imposes strong restrictions on it.
Using the very plausible form for $K$ conjectured in ref.[14], it is easy to
check that eq.(\ref{Wa}) still holds for $\delta^{GS}\neq 0$.}
(\ref{Wa}), which in this case reads
\begin{eqnarray}
A^{3} = \left(\frac{M}{3}\right)^{\frac{3N}{3N-M}}(32\pi^2 e)^
{\frac{3(M-N)}{3N-M}}\
\frac{e^{-24\pi^2kS/(3N-M)} } {[\eta(T)]^{6-(12k\delta^{GS}/(3N-M))}} \;\;,
\label{Ast2}
\end{eqnarray}
where $\delta^{GS}\equiv\sum_i\delta^{GS}_i$. Again, substituting
(\ref{Ast2}) in $W=W^{np}+W^{pert}$ gives a superpotential which
depends on $S$ and $T$ only
\begin{eqnarray}
W^{eff} & = & \tilde d\; \frac{e^{{-3kS}/{2\tilde{\beta}}} }
{[\eta(T)]^{6-(3k\delta^{GS}/4\pi^2\tilde\beta)}}\;\;\;,
\label{Weff2}
\end{eqnarray}
where $\tilde d$, $\tilde\beta$ are defined in eq.(\ref{dtilde}). Note
again that $W^{eff}$ has the correct modular weight $-3$ and a form
very similar to that of the pure YM case (\ref{super}). Using the K\"ahler
potential (\ref{K}), written for an overall modulus $T$, the scalar
potential for a generic $W(S,T)$ is
\begin{eqnarray}
V = \frac{1}{Y(2T_R)^3}
 \{  |YW_{S}-W|^{2} + \frac{Y}{3Y+\frac{1}{4\pi^2}
\delta^{GS}}|3W+\frac{1}{4\pi^2}\delta^{GS}W_S
%\nonumber \\
- 2T_RW_{T}|^2 - 3|W|^2 \}
\label{V2}
\end{eqnarray}
which is the same as (\ref{V}) when $\delta^{GS}_1=\delta^{GS}_2
=\delta^{GS}_3=\frac{1}{3}\delta^{GS}$. If this is not so, there
is a certain asymmetry between $T_1$, $T_2$ and $T_3$. Then the
results obtained from (\ref{V2}) for $T$ can be considered as
an average of $T_i$. In any case we will see shortly that the
effect of $\delta^{GS}\neq 0$ is normally small.

As was mentioned in subsection 2.5 the complexity of (\ref{V2})
requires the various scenarios to be analyzed numerically. It is
remarkable however that the results of this analysis show that,
whenever $S_R$ lies in the realistic range ($S_R\sim O(1)$), the
actual value of $S_R$ is still extremely close to that obtained
from condition (\ref{mincon1}). In particular, for a unique condensate
there is no minimum at all in $S_R$ and the shape of the potential
is as in fig.1. For two condensates
\begin{eqnarray}
W^{eff} = \tilde{d}_{1} \frac{e^{-3k_1S/2 \tilde\beta_1}}
{  [\eta(T)]^{ 6-(3k_1\delta^{GS}/4\pi^2\tilde\beta_1) } } +
\tilde{d}_{2} \frac{e^{-3k_2S/2 \tilde\beta_2}}
{  [\eta(T)]^{ 6-(3k_2\delta^{GS}/4\pi^2\tilde\beta_2) } }\;\;\;,
\label{supeff2}
\end{eqnarray}
condition (\ref{mincon1}) can be fulfilled and there appear lots of
interesting scenarios, as in the $\delta^{GS}=0$ case. Therefore
the corresponding value of $S$ is close to
\begin{eqnarray}
S_{R} & = &
S_{R}^{(\delta^{GS}=0)} \ +\
\frac{\delta^{GS}}{4\pi^2}\ \log |\eta(T)|^2
\nonumber \\
  & \simeq & \frac{2}{3( k_1\tilde{\beta}_{1}^{-1}- k_2\tilde{\beta}_{2}^{-1})}
\log\frac{\tilde{d}_1k_1\tilde{\beta}_1^{-1}}
{ \tilde{d}_2k_2\tilde{\beta}_2^{-1}}\ -\ \frac{0.645}{4\pi^2}\ \delta^{GS}
\label{sr4}\\
{\rm Im}\ S & = & {\rm Im}\ S^{(\delta^{GS}=0)}
\ +\ \frac{\delta^{GS}}{2\pi^2}{\rm phase}(\eta(T))\;\;,
\label{ims4}
\end{eqnarray}
where $S^{(\delta^{GS}=0)}$ is the value of $S$ for the $\delta^{GS}=0$
case, as given by eqs.(\ref{sr2},\ref{ims2}). In writing (\ref{sr4})
we have used the fact that $3S_Rk_a\beta_a^{-1}>>1$
(since $S_R\sim O(1)$ for viable scenarios)
and the empirical result that the
value of $T$ in the minimum obtained numerically continues being
$T\simeq 1.23$ in all cases (the discrepancies only appear
at the fourth significant figure). Consequently
\begin{equation}
\frac{Y}{2}=g_{string}^{-2}=S_R+\frac{1}{8\pi^2}\delta^{GS}\log(2T_R)
\simeq \frac{2}{3( k_1\tilde{\beta}_{1}^{-1}- k_2\tilde{\beta}_{2}^{-1})}
\log\frac{\tilde{d}_1k_1\tilde{\beta}_1^{-1}}
{ \tilde{d}_2k_2\tilde{\beta}_2^{-1}}\ -\ \frac{0.195}{4\pi^2}\ \delta^{GS} .
\label{Y2}
\end{equation}
Comparing with (\ref{sr2}), (\ref{ims2}) it is clear that if
$\delta^{GS}$ is not large (i.e. $\frac{1}{4\pi^2}
\delta^{GS}\stackrel{<}{\sim}1$, as it actually happens in all the calculations
 of
$\delta^{GS}$ performed up to now [17]) the results will not
differ basically from those for
$\delta^{GS}=0$. This fact is illustrated in
table 5 with a few examples and some typical values of $\delta^{GS}$.
We have also included the result for $\delta^{GS}=0$ to facilitate
the comparison. The shape of the potential for
a particular example is given in fig.3, to be compared with the
corresponding one for $\delta^{GS}=0$, fig.2.
On the other hand the gravitino mass for $\delta^{GS}\neq 0$ is
given by
\begin{eqnarray}
m_{3/2} = e^{K/2} |W^{eff}| & \simeq & \frac{1}{Y^{1/2}(2T_R)^{3/2}
|\eta(T)|^{6-[12\delta^{GS}/(3N-M)]}}\
|\tilde{d}_{1}(1-\frac{k_1\tilde\beta_2}{k_2\tilde\beta_1})|\
e^{-3k_1S_R/2 \tilde\beta_1}
\nonumber \\
 & \simeq &
\left(\frac{2S_R^{(\delta^{GS}=0)}}{Y}\right)^{1/2} m_{3/2}^{(\delta^{GS}=0)}
\;\;,
\label{m322}
\end{eqnarray}
where $m_{3/2}^{(\delta^{GS}=0)}$ is given in eq.(\ref{m32}) and
we have used the first equation of (\ref{sr4}), thanks to which
the dependence of $m_{3/2}$ on $\delta^{GS}$ has almost been cancelled
(there only remains a slight dependence through the value of $Y$).
This remarkable stability of $m_{3/2}$ with
respect to variations of $\delta^{GS}$ is illustrated in table 5.
All the numbers of table 5 have been obtained by
numerical minimization of (\ref{V2}) and by using the exact formula
for $m_{3/2}$, but they fit very well the approximate expressions
(\ref{Y2}, \ref{m322}). So we conclude that
the results obtained for the simplest $\delta^{GS}=0$ case are
essentially maintained for the general $\delta^{GS}\neq 0$ case.

So far we have considered only untwisted hidden matter. The
most general situation, however, involves twisted matter. In fact,
it can be argued on grounds of modular invariance that for
$Z_3$, $Z_7$ orbifolds there cannot be two gauge groups with
different $k^{-1}\beta$ coefficients and all the matter in the untwisted
sector.The argument is very similar to that expounded in subsection
2.5. If only untwisted matter is
present it is clear from (\ref{bp}) that $\frac{1}{16\pi^2}\sum b_i^{'a}=
\tilde\beta^a$, i.e.
the complete beta function. Now, for $Z_3$ and $Z_7$ orbifolds the
coefficient of $\log(\eta(T_i))$ in eq.(\ref{fw}) must vanish,
i.e. $b^{'a}_{i}=2k^a\delta^{GS}_i$. Taking
into account that $\delta^{GS}_i$ are universal for all the gauge group
factors, this cancellation cannot take place unless all the gauge groups
have the same
$(k^{-1}\tilde\beta)$ factors, a situation which is useless for the practical
purpose of fixing the dilaton since this is equivalent to having a unique
condensate. For the rest of $Z_N$ orbifolds the
restriction is not so strong, since all of them have an $N=2$ sector
(the $Z_6$--II has two) associated with a fixed torus. In consequence
there is a plane, say the $i=3$ plane, in which the previous cancellation
does not take place. This means, in particular, that if for a $SU(N)$
factor there are singlet fields $A_i$ ($i$ denotes a holomorphic index)
giving mass to $M_i$ quarks (so $\sum_i M_i=M$), then the quantities
$k^{-1}(N-M_1)$ and $k^{-1}(N-M_2)$ must be the same as in any other
$SU(N)$ factor. In principle, there would be no restrictions on
$M_3$.

Let us consider now gaugino condensation in the presence of twisted
matter. Following similar steps to those of subsection 3.1 we work
with an overall modulus $T$ and a generic $A$ field of the twisted
type giving mass to $M(N+\bar N)$ representations through a trilinear
superpotential $W^{tr}=\sum_{\alpha=1}^{M}h(T)AQ_{\alpha}\bar Q_{\alpha}$
with $h(T)\neq$ const. In consequence, $\det{\cal M}=[h(T)A]^M$ in
(\ref{cond}). Analogously, the perturbative piece of eq.(\ref{supert})
becomes now $T$--dependent, $W^{pert} = \hat{h}(T)A^{3}$. The $A$ field
can be integrated out again through condition (\ref{Wa}), leading now to
\begin{eqnarray}
A^{3} = \left(\frac{M}{3\hat h(T)}\right)^{\frac{3N}{3N-M}}
[h(T)]^{\frac{3M}{3N-M}}
(32\pi^2 e)^{\frac{3(M-N)}{3N-M}}\
\frac{e^{-24\pi^2kS/(3N-M)} }
{[\eta(T)]^{6(\sum_ib^{'}_i-2k\delta^{GS})/(3N-M)}} \;\;,
\label{Ast3}
\end{eqnarray}
which substituted in $W=W^{np}+W^{pert}$ gives a superpotential that
depends on $S$ and $T$ only
\begin{equation}
W^{eff} = \tilde d_T\; \frac{e^{{-3kS}/{2\tilde{\beta}}} }
{[\eta(T)]^{3(\sum_ib^{'}_i-2k\delta^{GS})/8\pi^2\tilde\beta}} \;\;\;,
\label{Weff3}
\end{equation}
\begin{equation}
\tilde {\beta} = \frac{3N-M}{16\pi^2}\;,\;\;
\tilde{d}_T = (\frac{M}{3}-N)(32\pi^2 e)^{\frac{3(M-N)}{3N-M}}\left(
\frac{M}{3\hat{h}(T)}\right)^{\frac{M}{3N-M}}
(h(T))^{\frac{3M}{3N-M}}\;\;.
\label{dtilde3}
\end{equation}
This is very similar to eq.(\ref{Weff2}) for the untwisted case.
In fact the similarity is even stronger since the Yukawa couplings
$h(T)$, $\hat{h}(T)$ must have modular weights $(-3-n_A-2n_Q)$,
$-3(1+n_A)$ respectively in order to preserve the correct modular
weight of $W$. Therefore
$h(T)\propto[\eta(T)]^{-2(3+n_A+2n_Q)}$, $\hat h(T)\propto
[\eta(T)]^{-6(1+n_A)}$ and, in consequence,
\begin{eqnarray}
W^{eff}\sim  \tilde d\; \frac{e^{{-3kS}/{2\tilde{\beta}}} }
{[\eta(T)]^{6-(3k\delta^{GS}/4\pi^2\tilde\beta)}}\;\;\;,
\label{Weff4}
\end{eqnarray}
where $\tilde d$ is defined in $(\ref{dtilde})$ and we have used
eq.(\ref{bp}). Comparing with (\ref{Weff2}) we see that it is not
likely that the presence of twisted matter substantially changes
the results obtained for the untwisted case. It is remarkable
how the Yukawa couplings generate the piece $\sim\eta(T)$ in the
superpotential, even if this is not present at the beginning,
as it occurs for $Z_3$ and $Z_7$ orbifolds. What is even more:
the presence of twisted matter guarantees that the whole set of
moduli appears in the final expression of $W^{eff}$. This comes from
a kind of completeness that exists between the moduli involved
in the threshold corrections to $f_W$ (see eq.(\ref{fw})) and
those involved in the Yukawa couplings [27]. More precisely, the
moduli appearing in the threshold corrections are those associated
with the size and shape of the fixed tori of the orbifold, whereas
the remaining ones always appear in the expression of the twisted
Yukawa couplings. For instance, in the $Z_3$ and $Z_7$ constructions
there are no threshold corrections at all, but {\em all} the twisted
Yukawa couplings have a non--trivial dependence on the whole set
of moduli.

\subsection{Other scenarios}

In all the phenomenologically interesting scenarios above described
the perturbative interactions of the $A$ field (apart from its coupling
to the quarks) play a crucial role. However, though less probable, it
is also possible that those interactions are absent or that their
contribution to the scalar potential is flat. Then the only non--trivial
part of the superpotential is the non--perturbative one (\ref{cond})
and one may wonder whether there still exist interesting scenarios
under these circumstances. For a unique condensate, as it was explained
in subsection 3.1, this is not the case. The answer turns out to be
afirmative, however,
if the gauge group contains three or more subsectors. To see this,
consider first a hidden sector $SU(N_1)\times SU(N_2)$
with matter content $M_1(N_1+\bar N_1)+M_2(N_2+\bar N_2)$
where the two condensates are "sharing" the same $A$ field
in (\ref{cond}), i.e.\footnote{We choose here for simplicity
$\delta^{GS}_i=0$, $k=1$, an overall modulus $T$ and untwisted matter fields.}
\begin{eqnarray}
W^{np} = -N_1(32 \pi^{2} e)^{\frac{M_1}{N_1}-1}
A^{\frac{M_1}{N_1}}\ \frac{e^{-\frac{8 \pi^2}{N_1}S}}{[\eta(T)]^{6-2M_1/N_1}}
\ -\ N_2(32 \pi^{2} e)^{\frac{M_2}{N_2}-1}
A^{\frac{M_2}{N_2}}\ \frac{e^{-\frac{8 \pi^2}{N_2}S}}{[\eta(T)]^{6-2M_2/N_2}}
\label{shar}
\end{eqnarray}
Then the equation (\ref{Wa}) (which, as it was argued in subsection 3.1,
gives the minimum in $A$) has non--trivial solution
\begin{eqnarray}
A = \left(-\frac{M_2}{M_1}\right)^{\frac{N_1N_2}{N_2M_1-N_1M_2}}
(32\pi^2 e)^{-1}[\eta(T)]^{-2}
e^{-8\pi^2\frac{N_1-N_2}{N_2M_1-N_1M_2} S }
\label{Ast4}
\end{eqnarray}
(notice that if the $A$ fields for the two subsectors are different
each other the equation (\ref{Wa}) has no solution, apart from the trivial
one). Substituting
(\ref{Ast4}) in $W^{np}$ leads to
\begin{equation}
W^{eff} = D\; \frac{e^{-\alpha S} }
{\eta^6(T)}\;\;\;,
\label{Weff5}
\end{equation}
\begin{equation}
{\alpha} = 8\pi^2\frac{M_1-M_2}{N_2M_1-N_1M_2}\;,\;\;
{D} = -\frac{N_2}{32\pi^2 e} \left( 1-\frac{N_1M_2}{N_2M_1} \right)
\left(\frac{-M_2}{M_1}\right)^{\frac{N_1M_2}{N_2M_1-N_1M_2}}
\label{D}
\end{equation}
The form of $W^{eff}$ is very similar to the pure YM case for
a {\em single} condensate, see eq.(\ref{super}). Consequently, the
scalar potential does not have minima in $S$.
However, if we had started with the two condensates of eq.(\ref{shar})
{\em plus} a third, pure YM, $SU(N_3)$ condensate, the
corresponding effective superpotential would simply be
\begin{eqnarray}
W^{eff} = D \frac{e^{-\alpha S}}{\eta^{6}(T)} + d_{3}
\frac{e^{\frac{-8\pi^2}{N_3}S}}{\eta^{6}(T)}
\label{Weff6}
\end{eqnarray}
where $D$, $\alpha$ are defined in (\ref{D}) and $d_3$ is defined in
(\ref{super}). Then, as it was shown in subsection 2.3, the associated
scalar potential develops minima at $2S_RW_S^{eff}-W^{eff}=0$ and
$T=1.23$ or any other value related to this by duality. It turns out
to be many possibilities leading to reasonable values for the dilaton
and the gravitino mass $m_{3/2}=e^{K/2}|W^{eff}|$. Requiring
$1\leq S_R\leq 3.5$, $10^2\ GeV\leq m_{3/2}\leq 10^4\ GeV$ we find
72 scenarios of this type, which are given in table 6. Four of
them satisfy $N_a>M_a$. Simplified
expressions for $S_R$, $m_{3/2}$ can be obtained from the approximate
condition $W^{eff}_S=0$ (see eq.(\ref{Ws}))
\begin{eqnarray}
S_R & \simeq & \frac{1}{\alpha - 8\pi^2/N_3}
\log\left|\frac{D\alpha}{d_3(8\pi^2/N_3)}\right|
%\left(\log \frac{M_1-M_2}{M_1}+\frac{M_2N_1}{N_2M_1-N_1M_2}
%\log\frac{M_2}{M_1} \right)
%\simeq \frac{1.3}{8\pi^2}
%\left(\frac{8\pi^2}{N_3}-\alpha\right)^{-1}
\label{Saprox3} \\
m_{3/2} & \simeq & \frac{1}{(2S_R)^{1/2}(2T_R)^{3/2}|\eta(T)|^6}\
\left|d_3\left(1-\frac{8\pi^2}{N_3\alpha}\right)\right|\
e^{\frac{-8\pi^2}{N_3}S}
%\nonumber \\
% & \simeq & \ 1.794\ |D\left(1-\frac{N_3\alpha}{8\pi^2}\right)|\
%\frac{e^{-\alpha S_R}}{(2S_R)^{1/2}}
\label{maprox3}
\end{eqnarray}
Using the previous expressions it can be seen that a realistic value
for $m_{3/2}$ requires $S_R\simeq 0.34\ N_3\simeq 0.32\
\frac{N_2M_1-N_1M_2}{M_1-M_2}$.

Similarly, for a hidden sector with four condensing groups
$G=\prod_{a=1}^4 SU(N_a)$, $\sum_{a=1}^4 M_a(N_a+\bar N_a)$ matter
representations and two different $A$ fields associated with $SU(N_1)\times
SU(N_2)$ and $SU(N_3)\times SU(N_4)$ respectively, the effective
superpotential is
\begin{eqnarray}
W^{eff} = D_1 \frac{e^{-\alpha_1 S}}{\eta^{6}(T)} +
D_2 \frac{e^{-\alpha_2 S}}{\eta^{6}(T)}
\label{Weff7}
\end{eqnarray}
with obvious notation. In this case the number of different combinations
giving $1\leq S_R\leq 3.5$, $10^2\ GeV\leq m_{3/2}\leq 10^4\ GeV$
is much higher. This comes from the fact that there is a larger
amount of parameters ($N_1,...,N_4;M_1,...M_4$) to play with.
More precisely, there are $2702$ cases from which
$40$ are displayed in table 7. Approximate expressions for $S_R$,
$m_{3/2}$ analogous to (\ref{Saprox3},\ref{maprox3}) can be
straightforwardly written. In this case a reasonable value
for $m_{3/2}$ requires $S_R\simeq 0.315\ \frac{N_2M_1-N_1M_2}{M_1-M_2}
\simeq 0.315\ \frac{N_4M_3-N_3M_4}{M_3-M_4}$.

The scenarios considered in this subsection are attractive in the
sense that they are independent from the details of the perturbative
superpotential for the singlet $A$ fields, provided this is flat,
even though this does not seem to be the most usual situation.

\section{Conclusions}

We have studied in a systematic and modular invariant way a
large class of string scenarios
for which gaugino condensation in the hidden sector provides
a potential source of supersymmetry breaking as well as a non--trivial
scalar potential for the dilaton $S$, whose real part corresponds
to the tree level gauge coupling constant (${\rm Re}\ S=S_R\sim
g_{gut}^{-2}$).
For the case of pure Yang--Mills condensation (i.e. no matter in the
hidden sector) we find that no realistic results can emerge, even if
the hidden sector gauge group $G$ is not simple: $G=\prod_a G_a$.
Things change dramatically in the presence of hidden matter, which
on the other hand, is the most frequent case. Then there arises a wide
and very interesting class of scenarios with two hidden
condensing groups $SU(N_1)\times SU(N_2)$ for which
the dilaton dynamically acquires a reasonable value ($S_R\sim 2$)
and supersymmetry is broken at the correct scale ($m_{3/2}\sim
10^3\ GeV$). There is no need for fine--tuning to obtain good values
for $S$ and $m_{3/2}$ simultaneously, actually it turns out that
for constructions with Kac--Moody level 1 they are correlated. We have
made an exhaustive classification of the working possibilities (see tables),
giving also very simple approximate formulae for $S_R$ and $m_{3/2}$
which are useful to select models. Other types of working scenarios with
three and four condensing groups have also been studied.

Remarkably, the results are basically independent from the value
of $\delta^{GS}$ (the contributions from the Green--Schwarz mechanism).
Furthermore,
the minimization of the scalar potential also gives a value
for the modulus $T$ (${\rm Re}\ T=R^2$, with $R$ the radius of the
compactified space) which in all cases is close to $T=1.23$. This
indicates that the theory is spontaneously compactified at this
very sensible scale, thus breaking target--space modular invariance
in a spontaneous way. Concerning this point, it is worth noticing
that Yukawa couplings induce a non--trivial dependence $\sim \eta(T)$
of the condensation superpotential on the whole set of moduli (see
subsection 3.2.),
even if this is absent at the beginning  (as it occurs for $Z_3$ and
$Z_7$ orbifolds, where there are no moduli--dependent threshold
corrections).

\vspace{0.3cm}
\noindent{\bf ACKNOWLEDGEMENTS}

\vspace{0.3cm}
We thank J. Louis for illuminating discussions. We also thank
L. Iba\~nez and D. L\"ust for very useful comments. The work of B.C.
was supported by a Comunidad de Madrid grant.

%\noindent{\bf REFERENCES}
%\vspace{0.3cm}

\newpage
\pagestyle{plain}
\pagenumbering{arabic}
\setcounter{page}{34}

\vspace{0.3cm}
\noindent{\bf FIGURE CAPTIONS}

\begin{description}

\item[Fig.1] Scalar potential $V$ versus $S_R={\rm Re}\ S$ for
a pure Yang--Mills $SU(8)_{k=1}$ hidden sector. ${\rm Im}\ S$ and
$T$ have been set at their minimizing values. For other
simple gauge groups the shape of the potential is similar.

\item[Fig.2] Scalar potential $V$ (in logarithmic units chosen for
convenience) versus $S_R={\rm Re}\ S=g_{string}^{-2}$,
$T_R={\rm Re}\ T$ for a $SU(6)_{k=1}\times SU(7)_{k=1}$ hidden sector with
$6(7+\bar 7)$ hidden matter representations and $\delta^{GS}=0$.
${\rm Im}\ S$ and ${\rm Im}\ T$ are set at their minimizing values. The
two minima (related each other by duality) correspond to $S_R=2.1576$;
$T_R=1.2345$, $0.8100$.

\item[Fig.3] The same example as in fig.2, but for $\delta^{GS}=20$.
$Y/2\equiv S_R + \frac{\delta^{GS}}{8\pi^2}\log(2T_R)=
g_{string}^{-2}$. The two minima correspond to $Y/2=2.0588$;
$T_R=1.2307$, $0.8125$.

\end{description}

\end{document}